\documentclass[twocolumn]{aastex63} 
\pdfoutput=1
\usepackage{CJKutf8}
\usepackage{pbox}
\usepackage{footnotebackref}

\submitjournal{ApJ}

\shorttitle{CGM constraints from SZ effect \& X-ray data}
\shortauthors{Xuanyi Wu et al.}
\graphicspath{{./}{plots/}}
\begin{document}

\makeatletter
\LetLtxMacro{\BHFN@Old@footnotemark}{\@footnotemark}

\renewcommand*{\@footnotemark}{%
    \refstepcounter{BackrefHyperFootnoteCounter}%
    \xdef\BackrefFootnoteTag{bhfn:\theBackrefHyperFootnoteCounter}%
    \label{\BackrefFootnoteTag}%
    \BHFN@Old@footnotemark
}
\makeatother

\title{Constraints on circum-galactic 
media from Sunyaev-Zel'dovich effects 
and X-ray data}

\correspondingauthor{Xuanyi Wu}
\email{xy-wu18@mails.tsinghua.edu.cn}
\author[0000-0002-0744-8022]{Xuanyi Wu}
\affiliation{Department of Astronomy, Tsinghua University, Beijing 100084, China}

\author[0000-0001-5356-2419]{Houjun Mo}
\affiliation{Department of Astronomy, University of Massachusetts, Amherst, MA01003, USA}

\author[0000-0002-8711-8970]{Cheng Li}
\affiliation{Department of Astronomy, Tsinghua University, Beijing 100084, China}

\author[0000-0001-6860-9064]{Seunghwan Lim}
\affiliation{Department of Physics and Astronomy, University of British Columbia, 
6224 Agricultural Road, Vancouver, BC V6T 1Z1, Canada
}

\begin{abstract}
We use observational measurements of thermal and kinetic 
Sunyaev-Zel'dovich effects, as well as soft X-ray emission associated 
with galaxy groups to constrain the gas density and  
temperature in the circumgalactic medium (CGM) for dark matter halos 
with masses above $10^{12.5}M_{\odot}$. A number of generic 
models are used together with a Bayesian scheme to make model 
inferences. We find that gas with a single temperature component 
cannot provide a consistent model to match the observational
data. A simple two-phase model assuming a hot component and an 
ionized warm component can accommodate all the three observations.
The total amount of the gas in individual halos is inferred to be 
comparable to the universal baryon fraction corresponding to the
halo mass. The inferred temperature of the hot component is comparable 
to the halo virial temperature. The fraction of the hot component
increases from $(15-40)\%$ for $10^{12.5}{\rm M}_\odot$ halos
to $(40-60)\%$ for $10^{14.5}{\rm M}_\odot$ halos, where the ranges 
reflect uncertainties in the assumed gas density profile.    
Our results suggest that a significant fraction of the halo gas 
is in a non-thermalized component with temperature much lower than 
the virial temperature. 
\end{abstract}


\section{Introduction}
\label{intro}

In the current paradigm of structure formation, galaxies are 
assumed to form and evolve in dark matter halos
\citep[see][for a review]{2010galaxyformation}. In this scenario,  
accretion and feedback together govern the growth of galaxies. 
The circum-galactic medium (hereafter CGM) is the repository for 
baryons, through which galaxies are connected to the intergalactic medium 
(IGM). Some of these baryons will be accreted by galaxies 
to form stars and central super-massive black holes, while most of them 
leave galaxies through outflows produced by supernova and quasars. 
Much of the ejected gas may be re-accreted, and so the gas may 
cycle through the CGM a number of times.  Currently, the evidence 
for this paradigm is indirect, mainly from investigations of 
galaxy stellar masses, metallicities, and star-formation rates 
over cosmic time. To study this paradigm directly, however,  
one needs to focus on the CGM, the ground zero for gas accretion 
and ejection.

The properties of the CGM has been studied in a number of ways. 
X-ray observations have been conducted to study the intra-cluster 
medium (ICM) in galaxy clusters and groups
\citep[e.g.][]{wangxray2014MNRAS.439..611W, anderson2015}. 
The results demonstrate that substantial amounts of gas heating 
is actually by non-gravitational sources, likely from the feedback 
of member galaxies \citep{Cavaliere_etal1998, ArnaudEvrard1999, HelsdonPonman2000,
KravtsovYepes2000,Babul_etal2002, McCarthy_etal2011,ashley2020}.
A complementary way to study the CGM is through the Sunyaev-Zel'dovich 
effect \citep[hereafter SZE;][]{SunyaevZeldovich1972} of free electrons 
on the spectrum of the cosmic microwave background (CMB) owing to 
inverse Compton scattering. The thermal SZE (hereafter tSZE) is 
proportional to the line-of-sight integral of the electron pressure 
(or thermal energy density), while the kinetic SZE (kSZE) is proportional 
to the integral of the momentum density along a given line
of sight, 
thus providing two independent constraints on the properties of 
the ionized part of the CGM.  Great efforts have been made to measure the SZE 
using CMB surveys, such as {\it Planck}\footnote{\url{https://www.nasa.gov/mission_pages/planck}}, the Atacama Cosmology Telescope  
({\it ACT})\footnote{\url{https://act.princeton.edu}}, and the South Pole Telescope ({\it SPT})\footnote{\url{https://pole.uchicago.edu}}. 
For example, using a sample of galaxy groups covering a large range of 
halo masses together with the {\it Planck} 
Compton parameter map \citep{Planck_XXI2014}, 
\citet{lim2018} found that the thermal content of the gas in 
low-mass halos is significantly lower than that expected from the 
simple self-similar model, where the halo gas is assumed to be 
at the virial temperature and to have a mass given by the universal
baryon fraction. By stacking galaxy groups with known masses and 
peculiar velocities in {\it Planck} maps, \citet{lim2020}
found that the total kSZE within halos implies a baryon to 
dark matter mass ratio that is comparable to the universal baryon 
fraction. The tSZE and kSZE results combined, therefore, indicate that not all 
the halo gas is at the virial temperature \citep{lim2020}.

Clearly, these observational results provide important information   
about the properties of the CGM. In this paper, we use the combination of 
SZE observations and soft X-ray data to constrain the density and 
temperature of the CGM. We use a set of generic models to show
what we can learn from the observational data. The paper is organized 
as follows. We describe the observational data used in our analysis 
in \S\ref{obser}, and models of halo gas in \S\ref{model}.
Our analysis and results are presented in \S\ref{result}.
Finally, we summarize and discuss our results in 
\S\ref{summary}. Throughout the paper, we adopt a flat universe with 
matter density $\Omega_{m}=0.308$ and the reduced Hubble 
constant $h=0.678$, as given in \citet{Collaboration2015Planck}.


\section{Observational data}
\label{obser}

\subsection{Thermal SZE}

The thermal Sunyaev-Zel'dovich effect (tSZE) is produced by the 
inverse Compton scattering of CMB photons as they encounter high energy thermal
electrons in hot gas \citep{SunyaevZeldovich1972,sze}. The tSZE 
presents itself as a distortion in the CMB spectrum, and its strength is 
characterized by the Compton $y$-parameter:
\begin{equation}\label{y_parameter}
y\equiv \frac{\sigma_T}{m_e c^2} \int P_e dl\,,
\end{equation}
where $\sigma_T$ is the Thompson cross-section,
$m_e$ is the electron mass, $P_e=k_B n_e T_e $ is the 
electron pressure with $k_{B}$ being the Boltzmann constant, 
$n_e$ the electron number density and $T_e$ the electron temperature.
The integration is along the line of sight (LOS).
For galaxy groups/halos concerned here, the total tSZE flux 
within a radius $R$ of an object is described by 
a quantity $Y_R$, defined as
\begin{equation}\label{Y_R}
Y_R=\frac{\sigma_T}{ m_e c^2}\frac{1}{d_A^2}
\int_{V_R} P_{e} dV\,,
\end{equation}
where $d_{A}$ is the angular diameter distance of the object, and 
$V_R$ is the volume within $R$. The intrinsic tSZE flux can then be 
expressed in terms of a normalized quantity,   
\begin{equation}\label{y_200}
{\tilde Y}_{R}\equiv E^{-2/3}(z) Y_{R}
\left(\frac{d_A(z)}{500{\rm Mpc}}\right)^{2},
\end{equation}
where 
\begin{equation}
E(z)=\frac{H(z)}{H_{0}} =\sqrt{\Omega_{m}(1+z)^{3}+\Omega_{\Lambda}}
\end{equation}
is included to make ${\tilde Y}_{R}$ independent of the 
redshift for a given halo mass. The intrinsic tSZE, therefore,  
provides a measure of the thermal energy of electrons.

In our analysis, we use the results obtained by \citet{lim2018},
who used a spatial filter to extract 
tSZE signals associated with galaxy groups of various halo masses
from the Planck NILC all sky tSZ Compton parameter map 
\citep{Remazeilles2011, PlanckCollaboration2016b}.
\citet{lim2018} assumed that the spatial pressure profile
follows the universal pressure profile (UPP) model of \citet{Arnaud2009The}.
They divided galaxy groups into a number of sub-samples according to 
halo mass and obtained the average tSZE amplitude, ${\tilde Y}_{200}$, 
for galaxy groups contained in the same halo mass bin. The quantity 
${\tilde Y}_{200}$ is the same as ${\tilde Y}_{R_{200}}$, where 
$R_{200}$ is the radius within which the mean density of 
the halo is 200 times the critical density of the universe. 
We define the halo mass $M_{200}$ as the mass of a halo within $R_{200}$. 
Since the Planck observation does not resolve the flux distribution 
for all individual halos, the values of ${\tilde Y}_{200}$ they obtained 
should be used in combination with the adopted UPP to get the total
tSZE flux of the halo. For our analysis, we convert the observational 
quantity to 
\begin{equation}\label{p200}
P_{200}\equiv \int_{V_{200}} P_{e} dV\,. 
\end{equation}
The black data points in the first panel of \autoref{fig:compare} show 
$P_{200}$ as a function of $M_{200}$, while the 
shadow region represents the error of the measurement.
Here $M_{200}$ is the halo mass defined by the dark 
matter mass enclosed by $R_{200}$.
	The error in the original data was obtained using 
the variance among different independent samples. 
There may be covariance among different halo mass bins. 
However, the correlation coefficients were found to be much 
smaller than one, and we thus ignore the covariance in our analysis.

\subsection{Kinetic SZE}

The kSZE is produced by Doppler's effect of CMB photons as 
they are scattered by electrons associated with galaxy systems  
that have bulk motion \citep{Sunyaev1970Small}:
\begin{equation}
k\equiv\left(\frac{\Delta T}{T_{\rm CMB}}\right)_{\rm kSZ}
=-\frac{\sigma_T}{c} \int n_e ({\bf v}\cdot{\hat{\bf r}})dl,
\end{equation}
where ${\bf v}$ represents the bulk motion peculiar 
velocity of the gas, ${\hat{\bf r}}$ is the unit vector along 
the LOS in question, and the integration is along the LOS. 
For halos with known peculiar velocities, one can estimate the 
following quantity from the kSZE signal:
\begin{equation}
K_{200} \equiv \frac{1}{ d_A(z)^2}  \sigma_T \int_{V_{200}}n_e dV.
\end{equation}
From this, one can obtain the intrinsic kSZE flux,  
\begin{equation}
{\tilde K}_{200} \equiv K_{200}
\left(\frac{d_A(z)}{500{\rm Mpc}}\right)^{2}\,.
\end{equation}
We use the data of $\tilde K_{200}$ obtained by \citet{lim2020}
for our analysis. They applied spatial filters to the Planck 
100, 143, and 217GHz channel maps to extract kSZE signals of
galaxy groups of different masses simultaneously, 
assuming that $n_{e}$ follows the profile described in 
\citet{Plagge2010SUNYAEV}.

The amplitude of the profile
for each group is assumed to depend on its halo mass. 
Together with the peculiar velocities of individual groups 
given by \citet{wang2012}, \citet{lim2020} obtained 
the amplitude of the ionized gas profile as a function of halo mass
by matching the model maps with the observational ones. 
We use their result of ${\tilde K}_{200}$ as a function of halo 
mass, $M_{200}$, which is shown as the black data points in the 
second panel of \autoref{fig:compare}.
The shadow region is the error of the measurement. The error in the original data was estimated using 
independent samples and methods that accounts for 
systematic effects. The correlation coefficients between 
different halo mass bins were found to be much smaller than 
one and are ignored in our analysis. There might also be 
covariance between the kSZE and tSZE measurements. Unfortunately 
such covariance is not quantified.

	Turbulent motions of the gas within individual groups/clusters 
may contribute to kSZE signal, thereby affecting the interpretation 
of the observation. Unfortunately, our knowledge about such motions 
is still poor. Using gas simulations, \citet{ruan2013} found that 
the impact of the turbulence is less than ten percent compared 
with the peculiar velocity in the kSZE signal even for clusters that 
have undergone a recent merger. This uncertainty is much smaller
than that represented by the error bars of the data. 
In addition, the results used here are the averages over many 
groups/clusters of similar masses. We thus expect that the effects 
generated by turbulent motions within individual systems are further 
reduced in the results.

\subsection{X-ray luminosity}

X-ray emissions from galaxy groups/clusters are 
produced by the hot intra-cluster/group gas generated 
by bremsstrahlung. The X-ray luminosity depends 
on the temperature, density and metallicity of the gas and can 
be written as:
\begin{equation}\label{lx}
L(\nu_1,\nu_2)= \int_V \int_{\nu_1}^{\nu_2} 
n_e({\bf r})n_H({\bf r}) \epsilon_{X}(T,\nu, Z) dV d\nu,
\end{equation}
where $\nu_1$ and $\nu_2$ specify the frequency range, 
$n_H$ is the total hydrogen number density, 
and $\epsilon_{X}(T,\nu, Z)$ is the emissivity at the  
temperature $T$, frequency $\nu$ and metallicity $Z$. 
We use the Astrophysical Plasma Emission Code (\href{http://atomdb.org}{APEC}) 
to calculate the emissivity. The code includes physical processes 
such as bremsstrahlung, radiative recombination and two-photon radiation. 
The exact assumption of the metallicity does not have a significant 
impact on our results and we use $Z=0.05 Z_\odot$ here. 
For our analysis, we use the X-ray measurements from 
\citet{anderson2015}, who obtained the X-ray luminosity for halos 
of different masses by using the ROSAT All-Sky Survey (RASS). 
They extracted an average X-ray luminosity for locally brightest galaxies 
(LBGs) of a given stellar mass by stacking the X-ray images
around them, using the LBG sample of \citet{P13}. 

	The same stacking procedure was applied 
to groups/clusters in numerical simulations to validate their 
method and error estimates. Their using of LBGs ensures the 
systems selected are well isolated, so that the covariance 
in the X-ray luminosity between them is expected to be small.

For high-mass galaxies, they converted the galaxy stellar mass 
into a halo mass using the simulation calibrations 
presented in \citet{P13}. For low-mass systems, they used an 
abundance matching method to estimate $M_{500}$ from the 
stellar mass. Their final results are summarized as the X-ray 
luminosity, $L_{X,500}$, in the energy range between 0.5 
and 2 KeV within the radius of $R_{500}$. Here $R_{500}$ 
is the halo radius within which the mean density is 500 times 
the critical density of the universe, and $M_{500}$ is 
the dark matter mass enclosed by $R_{500}$. 
Their results for the $L_{X,500}$-$M_{200}$ relation 
are shown as the black data points in the third panel of 
\autoref{fig:compare}. Here we convert $M_{500}$ into $M_{200}$ 
assuming a NFW profile. We note that the $L_{X,500}$ as a 
function of halo mass obtained by \citet{anderson2015} 
is very similar to that obtained by 
\citet{wangxray2014MNRAS.439..611W}.

 	As one can see, the X-ray luminosity depends both on gas 
density and temperature. The dependence on gas density 
is strong, $\sim n^2$, while the temperature dependence is 
relatively weak, $\sim T^{1/2}$. The $n^2$-dependence 
implies that the X-ray luminosity depends not only on the 
mean gas density but also the density profile and clumpiness
of the gas distribution. The kSZE and tSZE measurements 
described above provide two additional constraints on these 
two quantities: the kSZE constrains the average gas density without 
depending on gas temperature and profile, while the tSZE
constrains the average of $nT$ within individual halos.  
Thus, the three measurements can be used to constrain  
three independent properties of the halo gas. In addition to the 
average gas density and mean temperature with halos,  
we will use the data to constrain the gas fraction in the hot 
component assuming a two-phase medium, and to constrain 
the gas density profile.

\section{Models of Halo Gas}
\label{model}

In this section, we describe our methods to model the density and temperature of gas 
in halos. We use the electron number density, $n_{e}$, to represent the 
density of the 
gas. We also assume that the gas temperature is the same as the electron temperature,
$T_e$. We examine a number of different cases with different assumptions of the
gas density and temperature. 

\subsection{Single phase models}

As our fiducial assumption for the gas density, we use the following profile:  
\begin{equation}\label{ne}
    n_e(M_{200}, r)= n_0(M_{200})\left[1+(r/r_c)^2\right]^{-3\beta/2},
\end{equation}
where $r_c=R_{200}/c$, with $c$ being the concentration of the halo, 
and $\beta=0.86$. The amplitude of the profile, $n_0$, is assumed to depend 
on $M_{200}$, and we use the $c$-$M_{200}$ relation given by 
\citet{Bullock2010Profiles} to compute $r_c$. 
This profile is adopted from \citet{Plagge2010SUNYAEV}
and is the same as that used in \citet{lim2018}.  
The first set of models we consider assume that all the 
halo gas is in a single phase with a given temperature profile, 
$T_h$. This set of models are denoted by $M1$. 

The simplest assumption in modeling the hot gas temperature in 
galaxy groups/clusters is that the gas is at the virial temperature,  
\begin{equation}\label{tvir}
T_{vir}=\mu\frac{G M_{vir} m_p}{2k_B r_{vir}}\,,
\end{equation}
where $M_{vir}$ and $r_{vir}$ are the virial mass and radius of 
the halo, respectively, $m_p$ is the proton mass, $\mu$ is 
the mean molecular weight, $G$ is the gravitational constant.
In our analysis we take $M_{vir}=M_{200}$ 
and $r_{vir}=R_{200}$.

We therefore consider a model, $M1-T_{vir}$, where the hot gas 
temperature is assumed to be uniform within halos and 
is equal to $T_{vir}$, and 
the gas density is given by equation (\ref{ne}) with $n_0(M_{200})$ constrained
by the observational data. 

More generally, we follow \citet{1.5Tvir2002ApJ...579..571L} 
and consider models in which the gas temperature profile is given by 
\begin{equation}\label{tprofile}
    T(M_{200}, r)= T_0 (M_{200})\left[1+1.5(r/R_{200})\right]^{-1.6},
\end{equation}
where $T_{0}(M_{200})$ is the amplitude of the profile which 
is to be constrained by observational data. 

In general, the two amplitudes, $n_0$ and $T_0$, can both depend 
on halo mass, $M_{200}$, and the dependence may not be deterministic. During the analyzing, we use $n_0$ in unit of $m^{-3}$ and $T_0$ in unit of K.
To allow variance in the relations, we use the following 
model for $n_0 (M_{200})$:
\begin{equation}
\log_{10}(n_0)={\cal N}(\mu_n, \sigma_n),
\end{equation}
where ${\cal N}(\mu,\sigma)$ is a normal distribution with mean $\mu$ 
and variance $\sigma$. We find that the observational data can be 
described by the following simple assumptions:  
\begin{equation}\label{mu_n}
\mu_n=\mu_{n0}+
\alpha_n \log_{10}\left[\frac{M_{200}}{3\times 10^{14}{\rm M}_{\odot}}\right],
\end{equation}
and 
\begin{equation}\label{sigma_n}
\sigma_n=\sigma_{n0}+
\beta_n  \log_{10}\left[\frac{M_{200}}{3\times 10^{14}{\rm M}_{\odot}}\right],
\end{equation}
where $\mu_{n0}$, $\alpha_n$, $\sigma_{n0}$ and $\beta_n$ are 
model parameters to be determined by the observational data. 
Similarly, for the gas temperature, we assume 
\begin{equation}
    \log_{10}(T_0)= {\cal N}(\mu_T,\sigma_T),
\end{equation}
with 
\begin{equation}
\label{mu_T}
  \mu_T=\mu_{T0}+
\alpha_T \log_{10}\left[\frac{M_{200}}{3\times 10^{14}{\rm M}_{\odot}}\right],
\end{equation}
and 
\begin{equation}
\label{sigma_T}
  \sigma_T=\sigma_{T0}+
   \beta_T\log_{10}\left[\frac{M_{200}}{3\times 10^{14}{\rm M}_{\odot}}\right].
\end{equation}
The model parameters, $\mu_{T0}$, $\alpha_T$, $\sigma_{T0}$ and $\beta_T$ 
are again to be constrained by observational data. 
We assume that the variances in the density and temperature 
are independent, so that they can affect the predicted mean 
only for $L_X$, but the predicted dispersion for all three observational quantities may be affected.     
Based on these assumptions on gas density and temperature, 
we consider models in which variances are allowed in $n_0$ or $T_0$, 
or in both, as listed in Table \ref{tab:models}.

\begin{table*}
\centering
\caption{Models of halo gas.}
\begin{tabular}{cccc}
\hline\hline
Model             & Gas phase & Variance in $T_0$ & Variance in $n_0$\\
\hline
M1-$T_{vir}$      & 1-phase   & No;~$T=T_{vir}$  & No \\
M1-no~scatter     & 1-phase   & No               & No \\
M1-$T_0$          & 1-phase   & Yes              & No \\
M1-$n_0$          & 1-phase   & No               & Yes\\
M1-$T_0,n_0$      & 1-phase   & Yes              & Yes\\
\hline    
M2-$T_{vir}$     & 2-phase    & No;~$T=T_{vir}$  & Yes\\
M2-no~scatter    & 2-phase    & No               & No\\
M2-$T_0$         & 2-phase    & Yes              & No\\
M2'-$T_0$         & 2-phase    & Yes             & No; a flatter $n_e$ profile\\
M2-$n_0$         & 2-phase    & No               & Yes\\
M2-$T_0,n_0$     & 2-phase    & Yes              & Yes\\
\hline    
\end{tabular}
\label{tab:models}
\end{table*}

\subsection{Two-phase models}

Because of processes like radiative cooling, feedback, 
thermodynamic and hydrodynamic instabilities, halo gas is expected 
to be a multi-phase medium consisting of gas components at different 
temperatures \citep[e.g. chapter 8,]{2010galaxyformation}.
Unfortunately, the details of such multi-phase media is poorly 
understood. Here we consider a simple case in which the 
halo gas is a two-phase medium, with a hot component at temperature 
$T_h$ and a warm component with temperature $T_w$, 
both assumed to be fully ionized. 
These two components are assumed to be in pressure equilibrium, 
so that at any given location
\begin{equation}
n_h T_h= n_w T_w\,,    
\end{equation}
where $n_h$ and $n_w$ are the local densities of the hot and 
warm components, respectively. We model the mass fraction 
in the hot component as 
\begin{equation}\label{model_fh}
f_h\equiv \frac{M_h}{ M_h+M_w }=f_{h0}\left(\frac{M_{200}}{M_0}\right)^{\alpha_f},
\end{equation}
where $M_h$ and $M_w$ are the total masses in the two components, 
respectively. The volume filling factors of the two components, 
$\phi_h$ and $\phi_w$, are defined as 
\begin{equation}
\phi_h \equiv \frac{V_h}{ V_h+V_w} = 1-\phi_w\,,
\end{equation}
so that
\begin{equation}
\frac{M_h}{M_w}
= \frac{n_h \phi_h}{ n_w \phi_w} = \frac{f_h}{1-f_h}\,.
\end{equation}
It can then be shown that
\begin{equation}\label{volumfraction}
\phi_h=\frac{T_hf_h}{{T_h\cdot f_h+T_w(1-f_h)}}\,.
\end{equation}
Finally, the total gas density and pressure can be written as
\begin{equation}\label{conserve_ne}
    n_e=n_h \phi_h+n_w \phi_w=
        n_h \left(\phi_h+\frac{T_h}{T_w}\phi_w\right)\,, 
\end{equation}
and 
\begin{equation}
    P_e = \frac{n_e T_h}{\phi_h +\phi_w T_h/T_w}\,.
\end{equation}
As one can see, once models are adopted for 
$n_e$, $T_h$, $T_w$ and $f_h$, one can make model
predictions for the kSZE through $n_e$ and for the tSZE through 
$P_e$. For the X-ray luminosity, we assume that the warm
component contributes little, so that it can be obtained 
through $n_h$ and $T_h$ in the volume occupied by the 
hot component specified by $\phi_h$. 

For all the two-phase models, listed as M2 in
Table \ref{tab:models}, we assume that 
$T_w=10^{5}{\rm K}$ at the halo center and follows the same 
profile as $T_h$. Since the volume occupied by the warm phase 
is in general much smaller than that occupied by the hot phase, 
the exact value assumed for $T_w$ is not important, provided 
that it is much smaller than $T_h$ and sufficiently high to ensure 
the ionization fraction is close to one. We also assume that
$f_h$ is described by equation (\ref{model_fh}), $n_e$ by 
equation (\ref{ne}), and $T_h$ is either equal to $T_{vir}$ 
(Model M2-$T_{vir}$) or described by equation (\ref{tprofile}) 
with $T_0$ being the temperature at halo center. Similar to the single 
phase models, variances in $n_0$ and $T_0$ are included 
in some of the two-phase models, as described in Table 
\ref{tab:models}. We adopt the 
$\beta$-model of equation (\ref{ne}) for the $n_{e}$ profile,
with the concentration parameter $c$ describing the core radius.
A higher $c$ leads to more concentrated distribution of the gas.  
As mentioned earlier, we use the model of \citet{Bullock2010Profiles}
for $c$ as a function of halo mass. To examine the impact of 
the assumed gas profile on our results, we also consider 
a new model, M2'-$T_0$, which is the same as M2-$T_0$ except 
that the concentration parameter is reduced by a factor of two. 

\section{Analysis and Results}
\label{result}

\begin{table}
\caption{
Prior ranges of the model parameters and
the relevant equations are listed. Posteriors 
for the M2-$T_0,n_0$ model are listed as an example. Details of the models can be found in \S\ref{model}.}
\centering
\begin{tabular}{c|c|c|l}
\hline
Name & Prior & Posterior &Relevant equation \\
\hline
$\alpha_{T}$ & [0.5,1.8] & 0.79$\pm$0.07  & \autoref{mu_T} \\
 \hline
$\mu_{T0}$ & [5.5,8.5] & 7.89$\pm$0.07  & \autoref{mu_T} \\
 \hline
 $\sigma_{T0}$&[0,0.2]& 0.08$\pm$0.06   & \autoref{sigma_T} \\
 \hline
 $\beta_{T}$&[-1,0] & -0.20$\pm$0.16  &\autoref{sigma_T} \\
 \hline
 $\alpha_{N}$ & [-0.4,0.4] & 0.17$\pm$0.07  & \autoref{mu_T} \\
 \hline
 $\mu_{n0}$ & [2.5,4.5]& 3.36$\pm$0.05   & \autoref{mu_n} \\
 \hline
 $\sigma_{n0}$& [0,0.2] & 0.06$\pm$0.04  & \autoref{sigma_n} \\
 \hline
 $\beta_{n}$ & [-1,0] & -0.31$\pm$0.19  & \autoref{sigma_n} \\
 \hline
 $\alpha_{f}$ & [0,0.6]& 0.20$\pm$0.06   & \autoref{model_fh} \\
 \hline
 $f_{h0}$ & [0.3,0.9] & 0.32$\pm$0.07  & \autoref{model_fh} \\
 \hline
\end{tabular}
\label{tab:parameters}
\end{table}

\begin{figure*}

\begin{minipage}[t]{0.3\linewidth}
\centering
\includegraphics[width=0.98\linewidth]{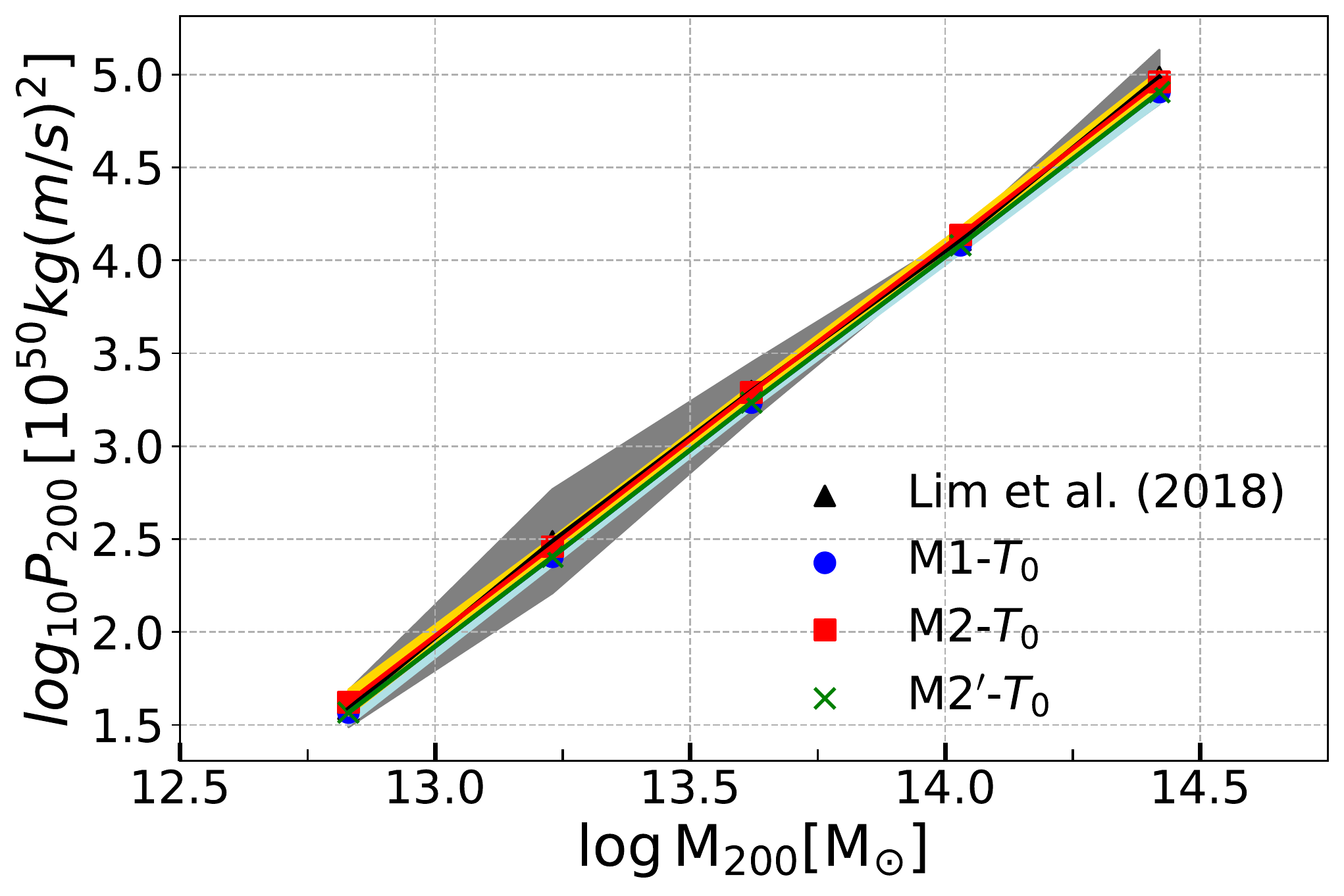}
\end{minipage}
\begin{minipage}[t]{0.3\linewidth}
\centering
\includegraphics[width=1\linewidth]{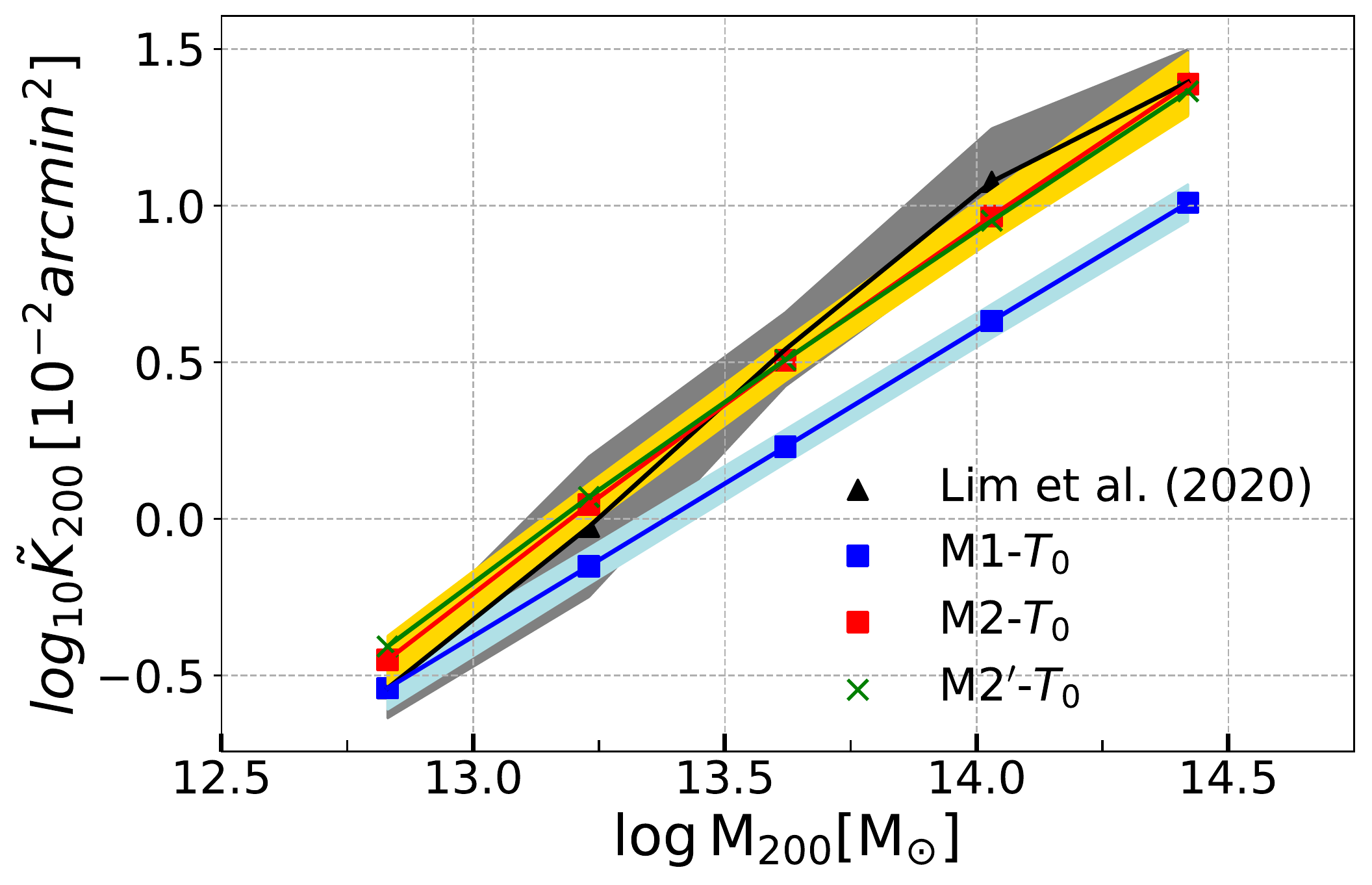}
\end{minipage}
\begin{minipage}[t]{0.3\linewidth}
\centering
\includegraphics[width=0.98\linewidth]{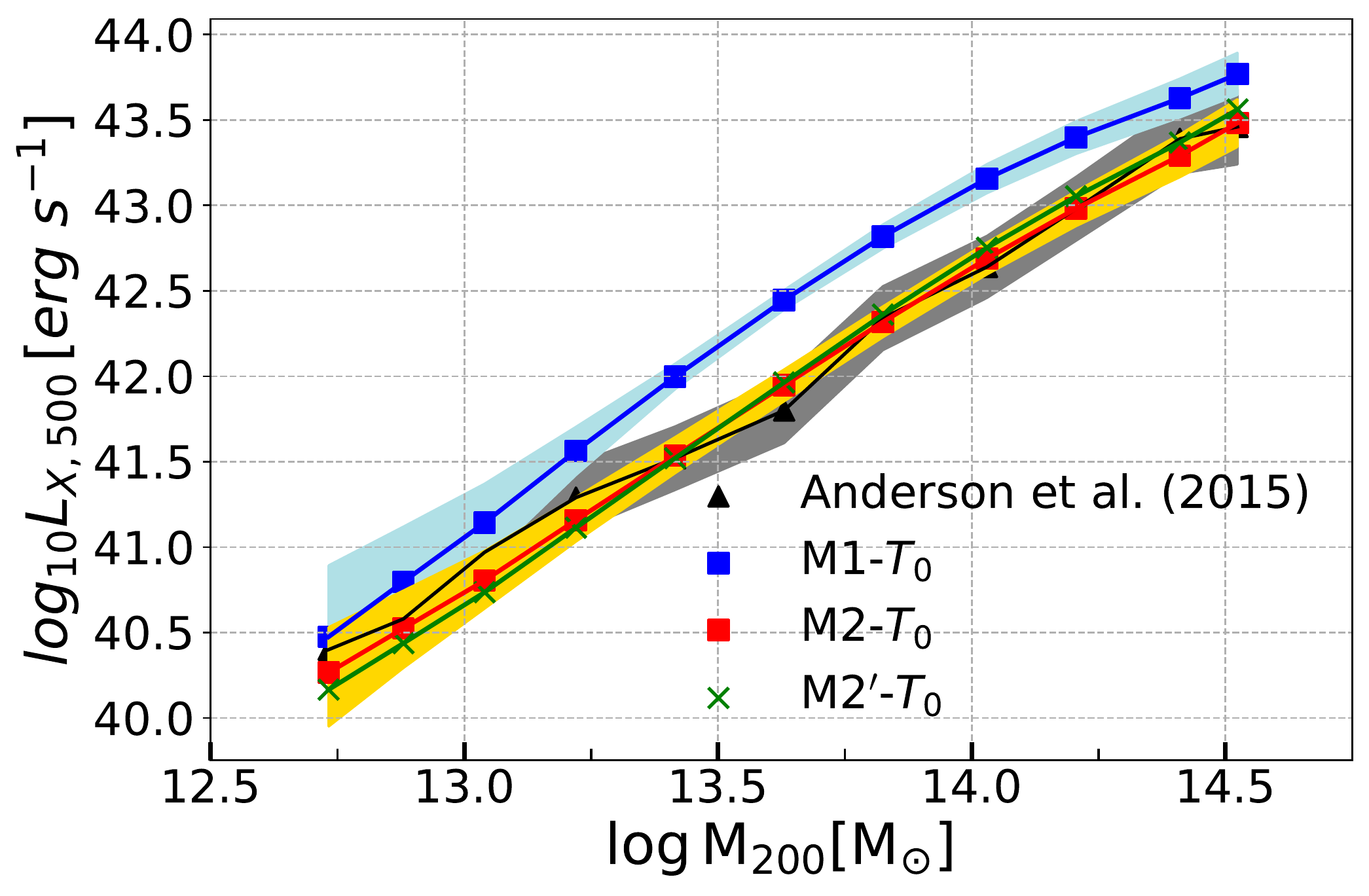}
\end{minipage}
\centering
\caption{Observational data used to constrain models. 
The black points and shadow regions are the mean and dispersion, 
respectively. The left, middle and right panels are, respectively,
the tSZE, kSZE and $L_X$, all plotted versus halo mass, $M_{200}$. 
For comparison, we plot the posterior predictions of three constrained 
models. Blue lines are for the single phase model, M1-$T_0$, red 
lines for the two-phase model, M2-$T_0$ and green line for M2'-$T_0$.
The shadow regions around the blue and red lines are the 90 
percentile ranges derived from the posterior distributions.}
\label{fig:compare}
\end{figure*}

\begin{figure*}
    \centering
    \includegraphics[width=0.8\linewidth]{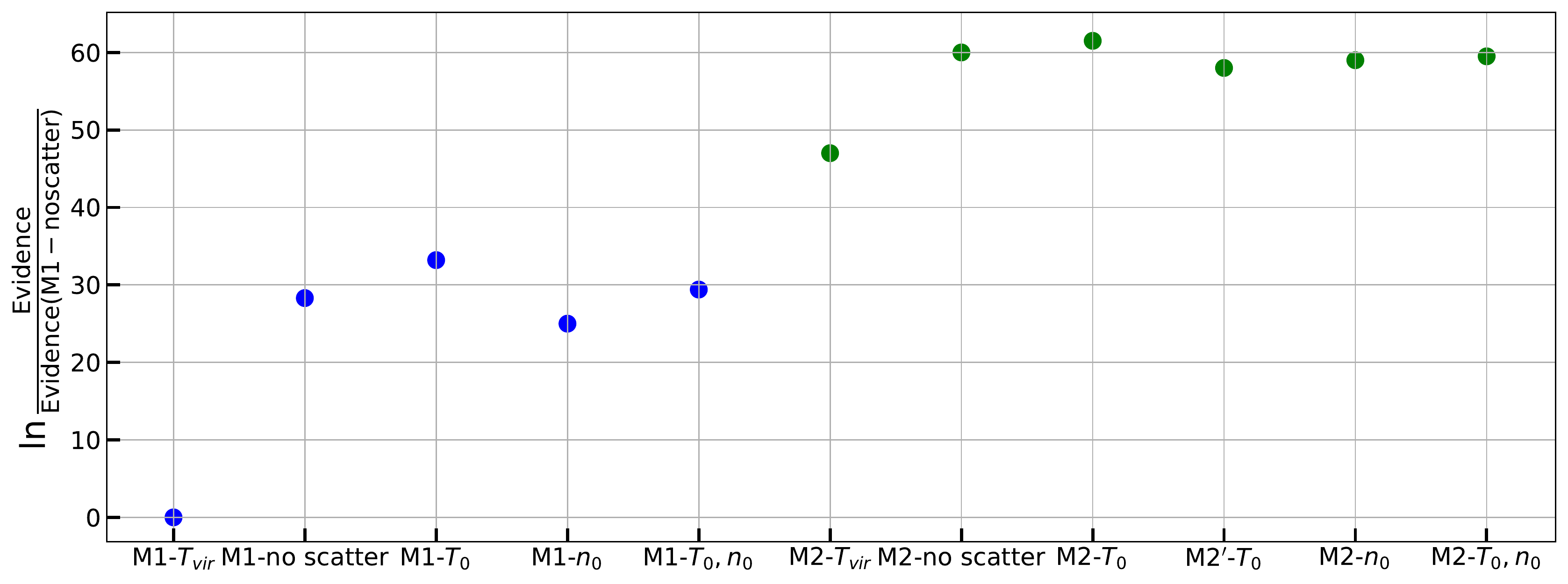}
    \caption{Bayesian evidences of different models with respect to M1-$T_{vir}$. 
    Blue points represent the set of single phase models (M1), while 
    green points are for the set of two-phase models (M2).}
    \label{fig:evidence}
\end{figure*}

\begin{figure*}
    \centering
    \includegraphics[width=0.9\linewidth]{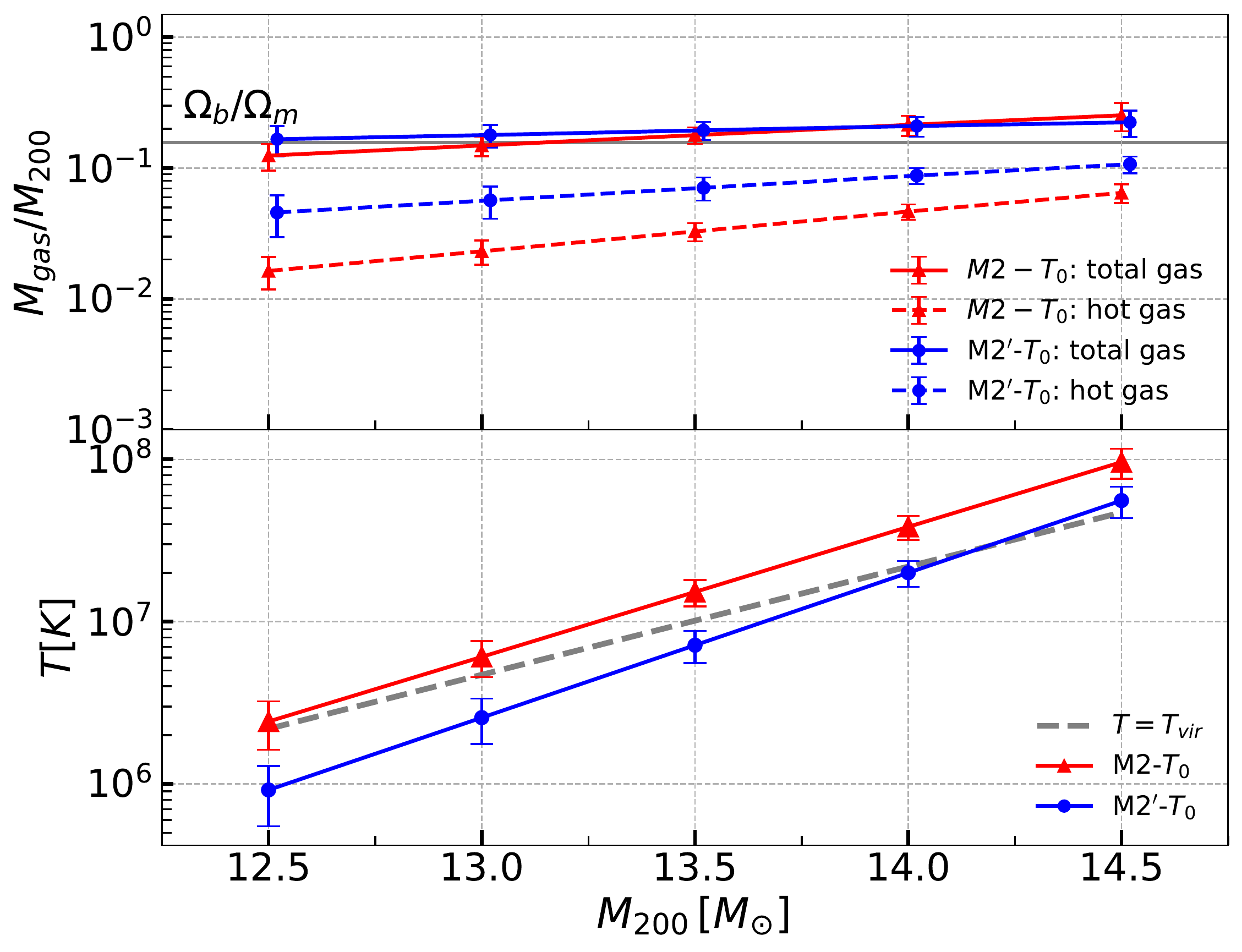}
    \caption{Top panel: the gas mass fractions predicted by
    M2-$T_0$ (red) and M2'-$T_{0}$ (blue). 
    Solid lines show the total gas mass fraction, 
    while dashed lines are for the hot gas mass fraction. 
    Error bars are the 90 percentile range of the predictions based on the 
    posterior distributions. Gray shadow region shows the cosmic baryon 
    fraction, $\Omega_b/\Omega_m=0.1574\pm0.002$, as given by 
    Planck cosmology\citep{Collaboration2015Planck}. The results of M2'-$T_0$ are slightly shifted 
    horizontally for clarity. Bottom panel: 
    hot gas temperature 
    as a function of $M_{200}$ predicted by $M2-T_{0}$ (red)
    and M2'-$T_0$ (blue). Error bars represents the 90 percentiles of 
    $T_{h}$ derived from the posterior distribution. 
    Gray dashed line is $T_{vir}(M_{200})$.}
    \label{fig:fb}
\end{figure*}

To constrain a model with observational data, we generate a set of 
$n_e$ and $T_h$ for a given set of model parameters.
We then compute the corresponding ${\tilde K}_{200}$, $P_{200}$ and $L_{X,500}$, 
and compare them with the observational data. For X-ray, we integrate 
the X-ray flux within $R_{500}$ to compare with observation.
The comparison between model predictions and observational data
is through the likelihood function,
\begin{equation}
	{\cal L} = \prod_{i} \frac{1}{\sqrt{2}\sigma_i}
	\exp\left[-\frac{({\cal M}_i (\Theta)-{\cal D}_i)^2}{ 2\sigma_i^2}\right],
\end{equation}
where ${\cal M}_i(\Theta)$ is the prediction of the model
specified by the parameter set $\Theta$ for the $i$th data point, 
${\cal D}_i$ is the corresponding observational data and
$\sigma_i$ is its error. The posterior distribution of model 
parameters is sampled with the PyMultinest, a Python module of the 
MultiNest sampling engine for both parameter estimation and 
model selection \citep{multinest}. We assume a flat
prior for each parameter, with its range chosen sufficiently 
broad (see Table~\ref{tab:parameters}). For reference, the posterior 
model parameters for M2-$T_0,n_0$ are also shown 
in Table~\ref{tab:parameters}; those for other models are given in the appendix.

To examine how different models can accommodate the observational data, 
we compute the Bayesian evidence,
\begin{equation}
{\cal E} ({\cal M})
=\int {\cal P} ({\cal D}\vert {\cal M}, \Theta) {\cal P}(\Theta\vert{\cal M}) d\Theta,
\end{equation}
where ${\cal P} ({\cal D}\vert {\cal M}, \Theta)$ is the probability of the 
data given model ${\cal M}$ with a given set of model parameters, 
$\Theta$, and ${\cal P}(\Theta\vert{\cal M})$ is the prior distribution of model 
parameters. So defined, the Bayesian evidence is the probability of the 
observational data in a model family ${\cal M}$, and thus can be used to
compare different model families in their abilities to accommodate the observational 
data. \autoref{fig:evidence} shows the logarithms (to natural base) of the Bayesian evidence ratio,
$\ln ({\cal E} ({\cal M})/ {\cal E} ({\cal M}_0))$, 
of a given model ${\cal M}$ relative to the reference model, 
${\cal M}_0$, which is chosen to be M1-$T_{vir}$. 
As one can see, all the two-phase models give a similar evidence ratio,
about 60 in logarithmic scales, while all the single-phase
models have a ratio about 30. Typically, a Bayesian 
evidence ratio of more than 20 in a logarithmic scale indicates 
a significant preference between the two models in comparison. It is thus 
clear that the combination of the tSZE, kSZE and $L_X$ data 
has a significant preference to the two-phase hypothesis (M2), while a
single phase model with gas at the virial temperature is the least favored. 

 To see more clearly how the models fit observational data,  
we use the posterior distribution of each individual model 
to predict the observations. As examples, the predictions 
of M1-$T_0$ and M2-$T_0$ are shown in \autoref{fig:compare}.
Clearly, M1-$T_0$ cannot accommodate the three sets of
observational data simultaneously: it significantly under-predicts
kSZE but over-predicts $L_X$. In contrast, M2-$T_0$ matches well all 
the three observations. We find that the results of these two models 
represent well those of the two model sets, M1 and M2, 
indicating that the key to match the observational data is the 
assumption of a two-phase medium, while other details, such as 
the inclusion of scatter in model parameters, only play a minor role.
We find that model M2-$T_{vir}$ matches the observational data 
only slightly worse than the other M2 models, suggesting that 
the average hot gas temperature needed to fit the data is not 
very different from $T_{vir}$. 
Model M2'-$T_0$ has a Bayesian evidence similar to 
M2-$T_0$, and their posterior predictions for the three sets of 
observations are also very similar, indicating that these observations 
do not provide a strong constraint on the gas density profile. 
However, as we will show below, the change of gas density profile
can affect the model inferences significantly.

We can use the posterior model distributions to make 
predictions for the gas mass and temperature as functions 
of halo mass. Here again we use M2-$T_0$ and M2'-$T_0$ 
as demonstrations. We estimate the baryon mass in halos
of a given mass by integrating the corresponding 
gas density profile within $R_{200}$, assuming a fully ionized 
primordial gas with a hydrogen mass fraction of $f_{H}=0.76$.
In the top panel of \autoref{fig:fb} we show the total
and hot gas masses, both normalized by halo mass,   
versus the halo mass. Red and blue lines are for  
M2-$T_0$ and M2'-$T_0$, respectively, while solid and 
dashed lines are for the total and hot gas, respectively.
For all cases, the error bars represent the 90 percentile of 
the posterior predictions. For comparison we also show    
the universal baryon fraction from \citet{Collaboration2015Planck}
as the horizontal gray band\footnote{The variance of universal baryon fraction is small here, so in the plot, the shadow region of baryon fraction just looks like a line.}. 

The posterior predictions of the central 
temperature of the hot gas component, $T_0(M_{200})$,  
by M2-$T_0$ and M2'-$T_0$ are shown in the bottom panel of 
\autoref{fig:fb}, together with the virial temperature, 
$T_{vir}(M_{200})$, given by equation (\ref{tvir}).  

The total gas mass fraction predicted is comparable to the universal 
baryon fraction, a result of the constraints mainly from the 
kSZE observation. The hot gas mass fraction predicted by M2'-$T_0$ is 
higher than that by M2-$T_0$, because X-ray emission is less 
efficient for the more extended gaseous halos in M2'-$T_0$.
In both cases, the hot gas mass fraction increases with 
$M_{200}$ with a logarithmic slope of about $0.2$, but is significantly 
lower than the universal fraction. For M2-$T_0$, the ratio 
between the hot gas fraction and the universal fraction increases
from $\sim 0.15$ to $\sim 0.4$ as halo mass changes from 
$10^{12.5}\,M_\odot$ to $10^{14.5}\,M_\odot$; for M2'-$T_0$,
the increases is from $\sim 0.4$ to $\sim 0.6$.
This suggests that a large fraction of the halo gas is
in an ionized phase with temperature much lower than the 
virial temperature. Because of the constraints from the tSZE, 
the hot gas temperature is required to be lower 
in M2'-$T_0$ than in M2-$T_0$ by a factor about two.
For both of these models, the central hot gas temperature, $T_0$,   
is comparable to the halo virial temperature within a factor of two. 
The predicted $T_0$-$M_{200}$ relations 
have a logarithmic slope of about 0.8, slightly steeper than 
that of the $T_{vir}$-$M_{200}$ relation. 

The differences in the posterior predictions
of the hot gas fraction and temperature between M2-$T_0$ 
and M2'-$T_0$ suggest that there is significant degeneracy
between the mass of hot gas halos and their profiles.  
Thus, without additional information about the profile, 
it is difficult to obtain stringent constraints on the
hot halo gas. The $\beta$-profile of equation (\ref{ne}) is 
motivated by observations of massive 
clusters \citep[e.g.][]{ne2017ApJ...843...28M}, and is roughly 
consistent with the hot gas profiles of clusters in gas 
simulations \citep{Lim_etal2020}. For halos of lower 
masses, the hot gas profile is not well known from observation.
In numerical simulations, the gas profiles of these lower-mass 
halos can be affected significantly by feedback. 
Using data from EAGLE \citep{eagleproject} and Illustris TNG 
\citep{nelson_tng1,fredrico_tng2,naiman_tng3,nelson_tng4,phillepich_tng5,volker_tng6}, \citet{Lim_etal2020} found that the $\beta$-model 
is a rough approximation to the halo gas in low-mass halos, 
provided that the core radius $r_c$ is increased by a factor of 
two. Thus, the results of models M2-$T_0$ and M2'-$T_0$ 
may represent the range expected from the uncertainties in 
the hot gas profile. However, if the core radius $r_c$ 
in equation (\ref{ne}) were comparable to the virial radius
for low-mass halos, so that the profile is approximately 
flat, the inferred mass in the hot component 
would approach the universal fraction, leaving no room for the 
presence of the warm component. In this case, the gas 
temperature would be about 10 times lower than the corresponding 
virial temperature, as shown in \cite{lim2020}.

\section{Summary and discussion}\label{summary}

In this work, we combine observational data of kSZE, tSZE and 
$L_{X}$ to constrain the density and temperature of
diffuse gas in halos. 
We use a number of generic models and a Bayesian scheme to 
explore the constraints provided by the observational data. 
Our main results can be summarized as follows. 
\begin{itemize}
\item 
Single phase models, in which all the halo gas is assumed 
to have similar temperature, cannot accommodate the observational 
data, suggesting that halo gas is not completely thermalized to a 
single phase. The tension may be alleviated if the gas density 
profile is much shallower than that seen in numerical simulations. 
A nearly flat profile is needed to explain the observational data 
with a single phase model.

\item
Simple two-phase models, which assume a hot gas component and 
an ionized warm component in pressure equilibrium, can match well with 
the observational data without depending on model details. 
\item 
The predicted total (hot plus warm) gas mass fraction in individual 
halos is comparable to the universal baryon fraction, suggesting 
that most halos can retain most of the baryons in their possession.
\item
The hot gas component in a halo has a temperature that is comparable 
to the virial temperature of the halo.
\item 
The fraction of the hot component is found to increase from 
$(15-40)\%$ for $10^{12.5}{\rm M}_\odot$ halos
to $(40-60)\%$ for $10^{14.5}{\rm M}_\odot$ halos, 
where the lower and upper bounds cover uncertainties in 
the assumed density profiles.
\end{itemize}

Our results have important implications for galaxy formation and evolution. 
Observations have shown that star formation in the universe is 
inefficient and that only a small fraction of the baryons is 
locked in stars \citep[e.g.][]{li2009}. To prevent baryons from forming stars too quickly, 
feedback processes are invoked either to heat the star forming gas or 
to eject it from galaxies. All numerical simulations of galaxy formation 
in the current paradigm need to incorporate some feedback processes 
to reproduce the stellar component observed in the universe. 
Most of the simulations seem to show the existence of hot gaseous 
halos with an average gas temperature comparable to the halo 
virial temperature and with mass significantly smaller than 
that implied by the universal baryon fraction 
\citep{Lim_etal2020}. These are consistent with 
our results for the hot gas component. However, in these simulations, 
a significant fraction of the baryons is ejected from halos
by feedback effects, so that the warm phase implied by our results 
is insignificant. This indicates that the current cosmological 
simulations may not be able to resolve multi-phase media or may have 
missed a significant non-thermalized gas component in gaseous halos.
Indeed, using high-resolution zoom-in simulations of clusters 
of galaxies, \citet{Nelson_etal2014ApJ792:25} found that a significant 
fraction of the gas pressure is non-thermal pressure produced
by the bulk motion of cooler gas. This is consistent with the results 
that a substantial fraction of the gas in clusters is in a phase 
with temperature much lower than the virial temperature. 
For low-mass halos, the presence of a gas component at temperature 
lower than the virial temperature has been probed by 
QSO absorption line systems such as MgII and OVI 
\citep[e.g.][]{werk2014,zhu2014,LanMo2018}. Unfortunately, the total 
amount of the gas implied by these systems are still uncertain
\citep[e.g.][]{Tumlinson2017}.

Clearly, with data from large, high-resolution CMB surveys, 
such as CMB-S4, and all sky X-ray surveys, such as eROSITA,
we hope to obtain much better constraints on the gaseous 
halos over a large mass range, so as to provide important insight 
about the processes by which gaseous halos and galaxies form and 
evolve.

\acknowledgments
This work is supported by the National Key R\&D Program of China
(grant No. 2018YFA0404502, 2018YFA0404503), and the National Science Foundation of China (grant Nos. 11821303, 11973030, 11673015, 11733004,
11761131004, 11761141012). SL acknowledges support by a CITA National Fellowship.

\bibliographystyle{aasjournal}
\bibliography{references}

\appendix
\section{Posteriors of two-phase models}

	 Here we list the posterior model parameters 
for all the two-phase models except M2-$T_{vir}$ and M2-$T_0$. 
The details of M2-$T_0$ are listed in 
\autoref{tab:parameters}, while M2-$T_{vir}$ cannot accommodate 
the three sets of observational data.

\begin{table}[h]
\caption{Posterior model parameters for some two-phase models.}
\centering
\begin{tabular}{c|c|c|c|c|l}
\hline
Parameter & M2-no~scatter & M2'-$T_0$ &M2-$n_0$ & M2-$T_0,n_0$ &discription\\
\hline
$\alpha_{T}$ & 0.81$\pm$0.09 & 0.96$\pm$0.07 &0.80$\pm$0.09 & 0.91$\pm$0.12 &\autoref{mu_T} \\
 \hline
$\mu_{T0}$ & 7.93$\pm$0.07& 7.43$\pm$0.10  & 7.96$\pm$0.08 & 7.98$\pm$0.08 &\autoref{mu_T} \\
 \hline
 $\sigma_{T0}$& - & 0.09$\pm$0.01 & - & 0.10$\pm$0.04 &\autoref{sigma_T} \\
 \hline
 $\beta_{T}$& -& -0.43$\pm$0.04  & - & -0.16$\pm$0.08 &\autoref{sigma_T} \\
 \hline
 $\alpha_{N}$ & 0.05$\pm$0.07 & -0.15$\pm$0.07& 0.06$\pm$0.07 & 0.14$\pm$0.11 &\autoref{mu_T} \\
 \hline
 $\mu_{n0}$ & 3.29$\pm$0.05 & 3.42$\pm$0.07 &  3.33$\pm$0.06 & 3.26$\pm$0.06 &\autoref{mu_n} \\
 \hline
 $\sigma_{n0}$& - & -  & 0.08$\pm$0.04 & 0.11$\pm$0.05  &\autoref{sigma_n} \\
 \hline
 $\beta_{n}$ & - & -  & -0.07$\pm$0.04 & -0.10$\pm$0.07 &\autoref{sigma_n} \\
 \hline
 $\alpha_{f}$ & 0.15$\pm$0.07& 0.10$\pm$0.07 & 0.14$\pm$0.07 & 0.14$\pm$0.06 &\autoref{model_fh} \\
 \hline
 $f_{h0}$ & 0.29$\pm$0.06 & 0.46$\pm$0.11 & 0.26$\pm$0.06 & 0.27$\pm$0.06&\autoref{model_fh} \\
 \hline
\end{tabular}
\label{tab:posteriors}
\end{table}
\end{document}